\documentclass[prd,twocolumn,superscriptaddress,floatfix,nofootinbib]{revtex4-2}
\pdfoutput=1 
\usepackage[T1]{fontenc}
\usepackage{amsmath}
\usepackage{amssymb}
\usepackage{amsthm}
\usepackage{amsfonts}
\usepackage{graphicx}
\usepackage{enumitem}
\usepackage{bm}
\usepackage{rotating}
\usepackage{hyperref}
\usepackage{pbox}
\usepackage[dvipsnames]{xcolor}
\usepackage{array}
\usepackage{physics}
\usepackage{dsfont}
\usepackage{mathtools}
\usepackage{leftidx}
\usepackage{lmodern}
\usepackage[normalem]{ulem}
\usepackage{braket}
\usepackage{tensor}
\usepackage{mathrsfs}
\usepackage{float}
\usepackage{dsfont}
\usepackage[margin=2cm]{geometry}
\usepackage[title]{appendix}
\usepackage{tcolorbox}

\usepackage{tikz}
\usetikzlibrary{arrows.meta,decorations.pathmorphing,math}

\renewcommand{\Re}{\mathrm{Re}}
\renewcommand{\Im}{\mathrm{Im}}
\renewcommand{\bar}{\overline}
\renewcommand{\tilde}{\widetilde}

\newcommand{\ii}{\mathsf{i}}

\newcommand{\sx}{\mathsf{x}}
\newcommand{\sy}{\mathsf{y}}

\newcommand{\bx}{{\bm{x}}}

\newcommand{\bk}{{\bm{k}}}

\newcommand{\rod}{\hat{\rho}_{\textsc{d}}^{0}}
\newcommand{\rao}{\hat{\rho}_{\textsc{a}}^{0}}

\newcommand{\rof}{\hat{\rho}_{\phi}^{0}}

\newcommand{\R}{\mathbb{R}}
\newcommand{\C}{\mathbb{C}}
\newcommand{\M}{\mathcal{M}}

\newcommand{\A}{\mathcal{A}}
\newcommand{\W}{\mathcal{W}}
\newcommand{\CS}{C^\infty_0(\M)}
\newcommand{\Sol}{\mathsf{Sol}}

\newcommand{\supp}{\text{supp}}

\begin{document}

\title{Non-perturbative simple-generated interactions with a quantum field for arbitrary Gaussian states}

\author{Erickson Tjoa}
\email{e2tjoa@uwaterloo.ca}
\affiliation{Department of Physics and Astronomy, University of Waterloo, Waterloo, Ontario, N2L 3G1, Canada}
\affiliation{Institute for Quantum Computing, University of Waterloo, Waterloo, Ontario, N2L 3G1, Canada}

\begin{abstract}
    In this work we first collect and generalize several existing non-perturbative models for the interaction between a single two-level qubit detector and a relativistic quantum scalar field in arbitrary curved spacetimes, where the time evolution is given by simple-generated unitaries, i.e., those generated by Schmidt rank-1 interaction Hamiltonians. We then extend the relativistic quantum channel associated to these non-perturbative models to include a very large class of Gaussian states of the quantum field, that includes an arbitrary combinations of coherent and squeezing operations (i.e., Gaussian operations) on the field. We show that all physical results involving the non-vacuum Gaussian states can be rephrased in terms of interaction with the vacuum state but with Gaussian operators applied to the field operators via the adjoint channel, effectively giving a ``Fourier transformed'' interpretation of the Gaussian operations in terms of the causal propagators in spacetime. Furthermore, we show that in these non-perturbative models it is possible to perform exact computation of the R\'enyi entropy and hence, via the \textit{replica trick}, the von Neumann entropy for the field state \textit{after the interaction} with the detector, without making any assumptions about the purity of the joint initial states of the detector and the field. This gives us a three-parameter family of ``generalized cat states'' of the field whose entropies are finite and exactly computable.
    

\end{abstract}

\maketitle

\section{Introduction}

In standard quantum information theory, the role of relativity is passive, in the sense that one accepts principles such as impossibility of superluminal signalling and derive consequences from them. Some of the no-go results such as Bell inequality and no-cloning theorem are intimately tied to the impossibility to send information faster than the speed of light. At the same time, it is well known a relativistic quantum theory requires us to work with quantum fields. In the absence of quantum theory of gravity, our best understanding of the interplay between quantum theory and gravity is given by the framework of quantum field theory (QFT) in curved spacetimes. Much of what is now known as \textit{relativistic quantum information} (RQI) seeks to understand various features and of QFT in curved spacetimes using the toolbox from quantum information theory. 

One of the most common and useful approaches in RQI involves the use of \textit{Unruh-DeWitt} (UDW) \textit{particle detector model} \cite{Unruh1979evaporation,DeWitt1979}, where one couples locally a qubit (which acts as a localized quantum-mechanical `detector') to a quantum field living on top of a generic curved spacetime. It is a simplified model of light-matter interaction representing a monopole-scalar model of atomic dipole-electromagnetic interaction in quantum optics. This model  has been refined to admit fully covariant description that allows for arbitrary trajectories and finite-size effect  \cite{Tales2020GRQO,Bruno2021broken}, as well as quantized centre of mass degrees of freedom \cite{Lopp2021deloc}, higher multipoles and spins. The UDW model is also useful for studying fundamental physics associated to relativistic trajectories or genuine quantum effects in curved spacetimes, such as the Unruh and Hawking effects. 

The more important advantages of the UDW model is that it is versatile  enough to provide answers to some fundamental questions that cannot be directly settled within quantum field theory in curved spacetimes. For example, it allows us to define local measurement theory \cite{polo2021detectorbased} for quantum fields even though projective measurements in quantum field theory violate relativistic causality \cite{sorkin1956}. Furthermore, since the UDW model is easily generalized to include multiple detectors, it is straightforward to apply it to study \textit{relativistic quantum communication} (RQC) between two localized parties in curved spacetimes \cite{Cliche2010channel,Jonsson2017quantum,jonsson2018transmitting,Simidzija2020transmit,Koji2020superadditive,Landulfo2016magnus1,Landulfo2021cost,tjoa2022channel}. There are numerous other applications of the UDW model in other contexts (see, e.g., \cite{pozas2015harvesting,pozas2016entanglement,tjoa2020harvesting,Tjoa2021notharvesting,smith2016topology,Aubry2018Vaidya,henderson2018harvesting,Gray2021imprint,Bruno2020neutrino,sahu2021sabotaging,Gallock2021nonperturbative} and references therein). 

Our work is largely motivated by the observation that there has been relatively few works on the characterization of relativistic quantum channels built using the UDW detector model. There has been some rather general and remarkable results, such as showing the entanglement-breaking nature of certain relativistic communication channels \cite{jonsson2018transmitting,Simidzija2020transmit} or proving ``no-go theorems'' on entanglement extraction from the quantum field \cite{Simidzija2018nogo}. However, these results are often restricted to Minkowski spacetime (which is nonetheless important) and relativistic quantum channels are much less understood compared to the non-relativistic counterparts. More recent work \cite{Landulfo2016magnus1,Landulfo2021cost,tjoa2022channel} has exploited the possibility of using non-pertubative methods to obtain very general results regarding certain class of relativistic quantum communication channels in arbitrary curved spacetimes, making use of the full power of algebraic approach to quantum field theory (AQFT). 

In this work we aim to fill this gap in the literature on three fronts. First, we will collect and unify some of the known non-perturbative coupling between the detector and the field: (i) the \textit{delta-coupling model}, (ii) the \textit{gapless model}, and (iii) the \textit{pure dephasing model}. Each speaks about different regimes: for example, (i) is about the short timescale of interaction relative to all other timescales, while (ii) and (iii) is about the internal dynamics of the qubit detector being much slower than other relevant timescales. We first show that these models are ``essentially'' equivalent as far as computation of physical observables are concerned, so we do not have to treat any of them separately until the very end when we need to crunch out specific numerical output (e.g., computing the actual matrix elements). Since they are all identified by the fact that the unitary evolution is generated by  Schmidt rank-1 Hamiltonians (analysed e.g., in  \cite{Simidzija2018nogo, Simidzija2020transmit} in the context of entanglement harvesting and communication), we follow these works and call it the family of \textit{simple-generated interactions}.

Secondly, these simple-generated interactions give rise to (relativistic) quantum channels such that the non-perturbative analysis straightforwardly extends to a very large class of Gaussian states of the field, including those that are not quasifree (vanishing one-point functions) such as the coherent states. These are states that can be constructed from a sequence of arbitrary displacement operators and also a large subclass of squeezing operators that are not ``momentum-entangled''. The idea is to embed these Gaussian operations into the Weyl algebra of observables for the QFT and convert calculations for Gaussian states into calculations involving vacuum states (which are straightforward) but with the field operators acted via the ``adjoint channel''. In effect, it is analogous to going into the Heisenberg picture and performing Gaussian operations on the observables instead. This seemingly trivial switch will give us quite a nice dividend: we will see that the calculations for arbitrary Gaussian states follow very straightforwardly from properties of the Weyl algebra and the Baker-Campbell-Hausdorff formula.

Finally, we show that in these non-perturbative models it is possible to perform exact computation of the R\'enyi entropy and the von Neumann entropy for the field state \textit{after the interaction} with the detector, without making any assumptions about the purity of the joint initial states of the detector and the field. This might appear surprising since most interactions introduce mixedness and there are very few states where the von Neumann entropy can be calculated explicitly. There are some exceptions: if the joint state of the detector and the field is initially pure, one can compute the resulting von Neumann entropy of the field by computing the von Neumann entropy of the detector using the fact that their joint unitary evolution maps pure states to pure states. Essentially, by representing the Gaussian operations as elements of the Weyl algebra, it is possible to compute the quantum entropies of the field \textit{algebraically}, and the von Neumann entropy can be computed using, for instance, the \textit{replica trick}. Therefore, the generalization of the non-perturbative channels to include arbitrary initial Gaussian states of the field gives us a way to compute the entropic quantities of a three-parameter family of ``generalized cat states'' of the field that are finite and exactly computable. To the best of our knowledge, most of these states do not admit simple path integral representations, thus we believe our calculations are of independent interests.

It should be stressed that by ``non-perturbative regime'' we mean that the unitary induced by the detector-field interaction for both gapless and delta-coupling approaches can be worked out without performing any truncation in the sense of Dyson series expansion. This relies on the fact that we can handle the time-ordering operation directly in these settings, so the unitary can be written as a finite linear combination of tensor products of bounded operators. We do \textit{not} mean that this calculation is non-perturbative in the sense that we solved \textit{exactly} for the full interacting theory of the detector-field system as a dynamical system. We will thus refer to the latter has having \textit{exact solution} for the dynamical system, while here we give a \textit{non-perturbative solution}, i.e., non-perturbative in the usual sense of having no truncation of any series expansion. Some readers may also prefer to interpret gapless and delta-coupled regimes as the regime where effectively we are performing ``resummation'' of the series expansion of the unitary evolution.


As part of the goal to make the use of algebraic approach more accessible, we have provided a more condensed version of the review of AQFT for scalar fields described in \cite{Tjoa2022fermi,tjoa2022modest,tjoa2022channel}. We also provide a very brief introduction to von Neumann algebras for the full algebra of observables, which are relevant to make sense of density matrices in AQFT and the computation of the field channel. For example, this helps us understand the computations of R\'enyi entropy in terms of type I von Neumann algebra of the algebra of observables, in contrast to the more well-known fact about local algebras.

Our paper is organized as follows. In Section~\ref{sec: AQFT} we introduce the bare minimum of AQFT approach needed in this work.  In Section~\ref{sec: UDW-model} we introduce the Unruh-DeWitt detector model and collect the non-perturbative models under the class of simple-generated interactions. In Section~\ref{sec: channels} we study the resulting qubit channel and the corresponding complementary channel, defined by tracing out the qubit instead of the field. In Section~\ref{sec: gaussian-operations} we show how we can embed Gaussian operations as elements of the Weyl algebra and how it relates to the standard calculations where the field state is typically taken to be the vacuum state. In Section~\ref{sec: extra-stuff} we show how it is possible to compute explicitly the R\'enyi entropy and von Neumann entropy of the field after the interaction with the detector. We adopt the units $c=\hbar=1$ and we use mostly-plus signature for the metric.


\section{Lightning review of AQFT}
\label{sec: AQFT}

In this section we review some aspects of algebraic framework of QFT. We will only cover the very bare minimum to understand or perform calculations outlined in this work to keep it self-contained while not unnecessarily burdening the discussion with background information. For AQFT part, this will be a more condensed version of the summary given in \cite{tjoa2022modest,Tjoa2022fermi}, which in turn are based on \cite{wald1994quantum,Dappiaggi2005rigorous-holo,Moretti2005BMS-invar,Khavkhine2015AQFT,fewster2019algebraic}.
Readers can also skip to Section~\ref{sec: UDW-model} onwards or Section~\ref{sec: channels} if they are more interested in the main content of this work, referring to this section only when certain details need to be consulted.

\subsection{Algebra of observables}

We consider a real scalar field $\phi$ in $(n+1)$-dimensional globally hyperbolic Lorentzian spacetime $(\mathcal{M},g_{ab})$. The field obeys the Klein-Gordon equation 
\begin{align}
     P\phi = 0\,,\quad  P = \nabla_a\nabla^a - m^2  - \xi R\,,
     \label{eq: KGE}
\end{align}
where $\xi \geq 0$, $R$ is the Ricci scalar and  $\nabla$ is the Levi-Civita connection with respect to $g_{ab}$. Global hyperbolicity means that $\M\cong \R\times \Sigma$ where $\Sigma$ is a Cauchy surface: in such spacetimes, the Klein-Gordon equation admits well-posed initial value problem throughout and we also have a good notion of ``constant-time slices''. For example, in flat space we have natural global coordinates $(t,\bx)$, with Cauchy surface $\Sigma\cong \R^n$ and any constant-$t$ surfaces serve as a good Cauchy surface.

In the large scheme of things, quantization in algebraic framework makes a great deal of use of ingredients in classical field theory. The idea is that we need to construct \textit{algebra of observables} $\A(\M)$ for the field theory as well as quantum states on which $\A(\M)$ acts.  We will see that the building blocks of the QFT come from constructing solutions of the wave equation \eqref{eq: KGE}. These solutions can be built using appropriate choice of Green's functions, and we need to provide a ``symplectic structure'' to realize the dynamical content of the theory, including the implementation of \textit{canonical commutation relations} (CCR). Finally, we need to construct quantum states without reference to any Hilbert space structure, due to the well-known existence of many unitary inequivalent Hilbert space representations. We will see that there are \textit{a priori} too many options, and the consensus is to pick a subclass of \textit{Hadamard states} which encode the notion that all states should look ``the same'' locally and as close to flat space QFT as possible.


Let $f\in \CS$ be a smooth compactly supported test function on $\M$. The \textit{retarded and advanced propagators} $E^\pm\equiv E^\pm(\sx,\sy)$ associated to the Klein-Gordon operator $P$ are Green's functions obeying
\begin{align}
    E^\pm f\equiv (E^\pm f)(\sx) \coloneqq \int \dd V'\, E^\pm (\sx,\sx')f(\sx') \,,
\end{align}
where here $\dd V' = \dd^4\sx'\sqrt{-g}$ is the invariant volume element. These solve the inhomogeneous wave equation $P(E^\pm f) = f$.  The \textit{causal propagator} is defined to be the advanced-minus-retarded propagator $E=E^--E^+$. The relevant fact for us is the following: if $O$ is an open neighbourhood of some Cauchy surface $\Sigma$ and $\varphi \in \Sol_\R(\M)$ is any real solution to Eq.~\eqref{eq: KGE} with compact Cauchy data, then there exists $f\in \CS$ with $\supp(f)\subset O$ such that $\varphi=Ef$ \cite{Khavkhine2015AQFT}.

In AQFT, the quantization of the real scalar field theory $\phi$ is to be viewed as an $\R$-linear mapping from the space of smooth compactly supported test functions to a unital $*$-algebra $\A(\M)$ given by
\begin{align}
    \hat\phi: C^\infty_0(\mathcal{M})&\to \A(\M)\,,\quad f\mapsto \hat\phi(f)\,,
\end{align}
which satisfy the following properties:
\begin{enumerate}[leftmargin=*,label=(\alph*)]
    \item (\textit{Hermiticity}) $\hat\phi(f)^\dag = \hat\phi(f)$ for all $f\in \CS$;
    \item (\textit{Klein-Gordon}) $\hat\phi(Pf) = 0$ for all $f\in \CS$;
    \item (\textit{Canonical commutation relations}  (CCR)) $[\hat\phi(f),\hat\phi(g)] = \ii E(f,g)\openone $ for all $f,g\in \CS$, where $E(f,g)$ is the smeared causal propagator
    \begin{align}
        E(f,g)\coloneqq \int \dd V f(\sx) (Eg)(\sx)\,.
    \end{align}
    \item (\textit{Time slice axiom})  $\A(\M)$ is generated by the unit element $\openone$ and the smeared field operators $\hat\phi(f)$ for all $f\in \CS$ with $\supp(f)\subset O$, where $O$ a fixed open neighbourhood of some Cauchy slice $\Sigma$.
\end{enumerate}
We say that $*$-algebra $\A(\M)$ is the \textit{algebra of observables} of the field. The \textit{smeared} field operator reads
\begin{align}\label{eq: ordinary smearing}
    \hat\phi(f) = \int \dd V\hat\phi(\sx)f(\sx)\,.
\end{align}
The (unsmeared) field operator $\hat\phi(\sx)$ commonly used in canonical quantization should be thought of as an operator-valued distribution. 

The dynamical content of the field theory is reflected by the symplectic structure as follows. The vector space of solutions $\Sol_\R(\M)$ can be equipped with a symplectic form $\sigma:\Sol_\R(\M)\times\Sol_\R(\M)\to \R$, defined as\footnote{As is well-known, this definition is independent of the choice of Cauchy surface.}
\begin{align}
    \sigma(\phi_1,\phi_2) \coloneqq \int_{\Sigma_t}\!\! {\dd\Sigma^a}\,\Bigr[\phi_{{1}}\nabla_a\phi_{{2}} - \phi_{{2}}\nabla_a\phi_{{1}}\Bigr]\,,
    \label{eq: symplectic form}
\end{align}
where $\dd \Sigma^a = -t^a \dd\Sigma$, $-t^a$ is the inward-directed unit normal to the Cauchy surface $\Sigma_t$, and $\dd\Sigma = \sqrt{h}\,\dd^n\bx$ is the induced volume form on $\Sigma_t$ \cite{Poisson:2009pwt,wald2010general}. The field operator $\hat\phi(f)$ can be expressed as \textit{symplectically smeared field operator}  \cite{wald1994quantum} 
\begin{align}
    \label{eq: symplectic smearing}
    {\hat\phi(f) \equiv \sigma(Ef,\hat\phi)\,,}
\end{align}
and the CCR algebra can be written as 
\begin{align}
    {[\sigma(Ef,\hat\phi),\sigma(Eg,\hat\phi)] = \ii\sigma(Ef,Eg)\openone = \ii E(f,g)\openone \,,}
\end{align}
where $\sigma(Ef,Eg) = E(f,g)$ in the second equality follows from Eq.~\eqref{eq: ordinary smearing} and \eqref{eq: symplectic smearing}. While in our case it is not directly necessary to construct $\A(\M)$ with explicit reference to $\sigma$, the symplectic form \eqref{eq: symplectic form} will be essential when we want to make connection to standard canonical quantization. In particular, we will need to define Klein-Gordon inner product for the one-particle Hilbert space associated to ``positive-frequency solutions''.

Since $\hat\phi(f)\in \A(\M)$ are unbounded operators, for free fields it is more convenient technically to work with its ``exponentiated version'' which forms a \textit{Weyl algebra} $\W(\M)$, whose elements are bounded operators. The Weyl algebra $\W(\M)$ is a unital $C^*$-algebra generated by elements that formally take the form 
\begin{align}
    W(Ef) \equiv 
    {e^{\ii\hat\phi(f)}}\,,\quad f\in \CS\,.
    \label{eq: Weyl-generator}
\end{align}
These elements satisfy \textit{Weyl relations}:
\begin{equation}
    \begin{aligned}
    W(Ef)^\dagger &= W(-Ef)\,,
    \quad 
    W(E (Pf) ) = \openone\,,\\
    W(Ef)W(Eg) &= e^{-\frac{\ii}{2}E(f,g)} W(E(f+g))
    \end{aligned}
    \label{eq: Weyl-relations}
\end{equation}
where $f,g\in \CS$. Note that  \textit{relativistic causality} (or \textit{microcausality}) is given by the third Weyl relations. For the rest of this work we try to stick mostly with $\W(\M)$.

\subsection{Algebraic states and quasifree states}

In AQFT the state is called an \textit{algebraic state}, defined by a $\C$-linear functional $\omega:\W(\M)\to \C$ (similarly for $\A(\M)$) such that 
\begin{align}
    \omega(\openone) = 1\,,\quad  \omega(A^\dagger A)\geq 0\quad \forall A\in \W(\M)\,.
    \label{eq: algebraic-state}
\end{align}
The state $\omega$ is pure if it cannot be written as $\omega= \alpha \omega_1 + (1-\alpha)\omega_2$ for any $\alpha\in (0,1)$ and any two algebraic states $\omega_1,\omega_2$; otherwise we say that the state is mixed. 

The relationship with standard canonical quantization comes from the \textit{Gelfand-Naimark-Segal (GNS) reconstruction theorem} \cite{wald1994quantum,Khavkhine2015AQFT,fewster2019algebraic}: we have a \textit{GNS triple} $(\mathcal{H}_\omega, \pi_\omega,{\ket{\Omega_\omega}})$, where $\pi_\omega: \mathcal{\W(\M)}\to {\mathcal{B}(\mathcal{H}_\omega)}$ is a Hilbert space representation with respect to state $\omega$. In its GNS representation, any algebraic state $\omega$ is realized as a \textit{vector state} $\ket{\Omega_\omega}\in\mathcal{H}_\omega$ and  $A\in \W(\M)$ are represented as bounded operators $\hat A\coloneqq \pi_\omega(A)\in \mathcal{B}(\mathcal{H}_\omega)$. We can thus write $ \omega(A) = \braket{\Omega_\omega|\hat A|\Omega_\omega}$. Since QFT in curved spacetimes has infinitely many unitarily inequivalent representations of the CCR algebra, the algebraic framework allows us to not pick any one of them until the very last step and work with all representations at once.

One of the most basic objects in QFT is the \textit{$n$-point correlation functions}\footnote{This is also known as Wightman $n$-point functions to distinguish it from other correlation functions.}, defined by
\begin{align}
    \mathsf{W}(f_1,...,f_n)\coloneqq \omega(\hat\phi(f_1)...\hat\phi(f_n))
    \label{eq: n-point-functions}
\end{align}
where $f_j\in \CS$ and for a fixed algebraic state $\omega$. It is to be understood that the RHS is computed within some GNS representation of $\A(\M)$. The GNS representation of the Weyl algebra $\W(\M)$ allows us to calculate Eq.~\eqref{eq: n-point-functions} by differentiation: for example,  the smeared Wightman two-point function reads
\begin{align}\label{eq: Wightman-formal-bulk}
    &\mathsf{W}(f,g) \equiv -\frac{\partial^2}{\partial s\partial t}\Bigg|_{s,t=0}\!\!\!\!\!\!\!\!\omega(e^{\ii\hat\phi(sf)}e^{\ii\hat\phi(tg)})
\end{align}
where the RHS is calculated in the GNS representation of $\W(\M)$ (since there is no good notion of derivatives directly on the Weyl algebra \cite{fewster2019algebraic}). As an example, in flat spacetime the vacuum GNS representation associated to vacuum state $\omega_0$ gives us the Minkowski vacuum $\ket{\Omega_{\omega_0}} = \ket{0_\textsc{M}}$.

The general agreement among AQFT practitioners is that physically reasonable states should be \textit{Hadamard states} \cite{KayWald1991theorems,Radzikowski1996microlocal}. Very roughly speaking, these states respect local flatness and finite expectation values of all observables appropriately \cite{KayWald1991theorems}. A particularly nice subclass of Hadamard states is \textit{quasifree states}: for these states, all odd-point functions in the sense of \eqref{eq: n-point-functions} vanish and all higher even-point functions can be written as in terms of just two-point functions\footnote{The term \textit{Gaussian states} refers to generalization when the one-point functions need not vanish and higher-point functions only depend on one- and two-point functions.}.  Well-known quasifree states are (squeezed) vacuum and thermal states; coherent states are non-quasifree Gaussian states.

At the end of the day, the reason why quasifree states are so useful and relevant is because it can be completely specified once we know the Wightman two-point functions associated to the quasifree state $\omega$: we have
\begin{align}
    \omega(W(Ef)) =   e^{-{\frac{1}{2}}\mathsf{W}(f,f)}\,.
    \label{eq: quasifree-definition}
\end{align}
At this point, we can simply take Eq.~\eqref{eq: quasifree-definition} as the \textit{definition} of quasifree states (see, e.g., \cite{tjoa2022modest,Tjoa2022fermi,KayWald1991theorems,Khavkhine2015AQFT} for more details). This is very useful because for most practical computations, we \textit{do} know how to calculate the smeared Wightman function especially if one is familiar with canonical quantization (many examples of the calculations can be found in standard texts such as \cite{birrell1984quantum}). 

The most important one is the vacuum state $\omega_0$, where we can write the (unsmeared) vacuum Wightman function as 
\begin{align}
    \mathsf{W}_0(\sx,\sy) &= \int \dd^n\bk\, u^{\phantom{*}}_\bk(\sx) u^*_\bk(\sy)\,,
\end{align}
where $u_\bk(\sx)$ are ``positive-frequency'' modes of Klein-Gordon operator $P$ normalized with respect to Klein-Gordon inner product $(\phi_1,\phi_2)_\textsc{kg}\coloneqq \ii\sigma(\phi_1^*,\phi_2)$, where $\phi_j\in \Sol_\C(\M)$ are complexified solutions to Eq.~\eqref{eq: KGE} (compare this with canonical quantization discussed in \cite{birrell1984quantum}). In situations where $\{u_\bk\}$ are known explicitly, we can often calculate the symmetrically smeared two-point function (sometimes exactly):
\begin{align}
    \mathsf{W}_0(f,f) = \int \dd V\,\dd V' f(\sx)f(\sy)\mathsf{W}_0(\sx,\sy)\,.
    \label{eq: Wightman-double-smeared}
\end{align}
In principle, we can compute any Wightman $n$-point functions for any algebraic state in their GNS representation. However, it is often most convenient to obtain the expression in relation to the vacuum representation, so that they take the form
\begin{align}
    \mathsf{W}(f,g) &= \mathsf{W}_0(f,g) + \Delta\mathsf{W}(f,g)\,,
\end{align}
where $\Delta\mathsf{W}(f,g)$ accounts for deviations from vacuum Wightman function  \cite{DeWitt1960radiation,wald1994quantum,KayWald1991theorems}. Some explicit calculations of $\Delta\mathsf{W}(\sx,\sy)$ in flat spacetime for Fock states, thermal states, coherent and squeezed states, can be found in \cite{simidzija2018harvesting,Tjoa2020resonance}, among many others.

\section{Covariant UDW detector model}
\label{sec: UDW-model}

Let us first review the covariant generalization of the Unruh-DeWitt (UDW) detector model that was developed in \cite{Tales2020GRQO,Bruno2021broken}. The detector is taken to be a two-level system with free Hamiltonian given by
\begin{align}
    \mathfrak{h}_0 &= \frac{\Omega}{2}(\hat\sigma^z+\openone)\,,
    \label{eq: free-hamiltonian}
\end{align}
where $\hat{\sigma}^z$ is the usual Pauli-$Z$ operator, whose ground and excited states $\ket{g},\ket{e}$ have energy $0,\Omega$ respectively. Let $\tau$ be the proper time of the detector whose centre of mass travels along the worldline $\sx(\tau)$.  The covariant generalization of the Unruh-DeWitt model can be defined by the interaction Hamiltonian density 
\cite{Tales2020GRQO,Bruno2021broken}
\begin{align}
    \hat{h}_I(\sx) &= f(\sx)\hat\mu(\tau(\sx))\otimes\hat{\phi}(\sx)\,.
\end{align}
Here $f\in \CS$ is the spacetime smearing function that prescribes the interaction region between the detector and the field in spacetime. 

On physical grounds, in the center-of-mass rest frame of the qubit detector, we should be able to separate the ``switching function'' that governs the duration of the interaction and the ``spatial profile'' of the detector (which would correspond to, say, atomic orbitals of a hydrogen atom). The coordinate system adapted to the center-of-mass trajectory of the qubit is the \textit{Fermi normal coordinates} \cite{Tales2022FNC} $\bar\sx=(\tau,\bar\bx)$, where the spacetime smearing function $f$ is factorizable into
\begin{align}
    f(\sx(\bar\sx)) \coloneqq \lambda\chi(\tau)F(\bar\bx)\,,
\end{align}
where $\lambda$ is the coupling strength, $\chi(\tau)$ is the (dimensionless) switching function, and $F(\bar\bx)$ is the spatial profile. The interaction unitary is given by (in the interaction picture) \cite{Tales2020GRQO}
\begin{align}
    \hat{U} = \mathcal{T}_\tau\exp\left[-\ii \int\dd V f(\sx)\hat\mu(\tau(\sx))\otimes\hat{\phi}(\sx)\right]\,,
\end{align}
where $\dd V = \sqrt{-g}\,\dd^n\sx$ is the invariant volume element. 

At this point, we may proceed to evaluate the time evolution perturbatively or non-perturbatively. There is a great deal of flexibility when one chooses to work within perturbative regime,  but there is mild causality violation and ``broken covariance'' whose origin can be traced to the combination of time-ordering and non-relativistic nature of the detector model \cite{Bruno2021broken}. In contrast, the non-perturbative methods allow us to probe beyond weak-coupling regime, though at the expense of a restricted types of dynamics where concrete calculations can be done. 

In this work, we are interested in unifying the non-perturbative computations and extract non-perturbative backreaction to the field due to the coupling between the detector and the field. This would correspond to two different types of regimes:
\begin{enumerate}[label=(\Alph*),leftmargin=*]
    \item \textbf{Delta-coupled detector regime}, where the interaction occurs at very short timescale, effectively at a single instant in time;
    
    \item \textbf{Gapless detector regime}, where the detector's energy level is taken to be degenerate (i.e., $\mathfrak{h}_0=0$); 
    
    \item \textbf{Pure dephasing regime}, where the interaction Hamiltonian density $\mathfrak{h}_I$ commutes with the free Hamiltonian (i.e., $[\mathfrak{h}_0,\hat{h}_I]=0$). 
\end{enumerate}

We will see that up to the choice of monopole operators and the spacetime smearing, these three regimes are in fact equivalent in the sense that the induced quantum channel has Kraus representation that is identical up to the choice of the monopole operators of the detector and spacetime smearing functions. Consequently, each of them have very different physical interpretations that we will describe later.

\subsection{Delta coupling}

The delta-coupled detector is the regime where the interaction timescale is assumed to be much faster than all the relevant timescales of the problem, so that the interaction can be taken to occur at a single instant in time (with respect to some time function, typically the detector's proper time or the global time function). This is often suitable to model one-shot fast processes instead of long-time processes such as thermalization and relaxation. If we assume that the detector interacts with the field only at $\tau=\tau_0$ in its own centre-of-mass rest frame, the spacetime smearing is given by
\begin{align}
    \mathfrak{f}(\sx(\bar\sx)) \coloneqq  \lambda\eta\delta(\tau-\tau_0)F(\bx)\label{eq: delta-smearing}\,.
\end{align}
One important caveat is that strictly speaking $\mathfrak{f}\not\in\CS$, so we should think of this delta-smearing as an appropriate limit of a sequence of compactly supported functions with decreasing width in the $\tau$-direction.

The unitary time evolution in the delta-coupling model simplifies greatly because the time ordering $\mathcal{T}_\tau$ is removed automatically --- there is nothing to time-order for a single-time interactions. That is, the unitary reduces to a simple exponential 
\begin{align}
    \hat{U} &= \exp\left[ -\ii \hat\mu(\tau_0) \otimes \hat\phi(\mathfrak{f}) \right]\,,
    \label{eq: unitary-delta-0}
\end{align}
where $\hat{\phi}(\mathfrak{f})$ is the smeared field operator with respect to the delta smearing \eqref{eq: delta-smearing}. Note that different $\tau_0$ simply labels different families of ``rotated'' monopole operators $\hat{\mu}(\tau_0)$, since
\begin{align}
    \hat{\mu}(\tau_0) &= \cos(\Omega\tau_0)\hat{\sigma}^x - \ii\sin(\Omega\tau_0)\hat{\sigma}^y\,.
\end{align}
Therefore, for each $\Omega\tau_0\in [0,2\pi)$ we have a one-parameter family of monopole operators spanned by the Pauli-$X$ and Pauli-$Y$ operators (since we fix the free Hamiltonian to be given by Pauli-$Z$ operator).

\subsection{Gapless detector}

The gapless detector regime is obtained by setting the energy gap in the free Hamiltonian $\Omega=0$, or equivalently we set the free Hamiltonian $\mathfrak{h}_0 = 0$. This is the regime where the internal dynamics of the detector is assumed to be much slower than all the relevant timescales of the problem, hence its internal dynamics are effectively frozen. Under this assumption, the expression simplifies greatly as the monopole operator is constant in time: $\hat{\mu}(\tau)=\hat{\mu}(0)\equiv \hat{\mu} $ for all $\tau$. The unitary operator then reduces to \cite{Landulfo2016magnus1}
\begin{align}
    \hat{U} = \mathcal{T}_\tau \exp\left[ -\ii \hat\mu \otimes \hat\phi(f) \right]\,,
    \label{eq: unitary-gapless}
\end{align}
where $\hat\phi(f)$ is the smeared operator for $f\in\CS$.  
The unitary operator can be evaluated non-perturbatively but we need to pass to Magnus expansion of $\hat{U}$, given by:
\begin{align}
    \hat{U} &= \exp\sum_{j=1}^\infty \Xi_j\,,
\end{align}
where 
\begin{equation}    
    \begin{aligned}
        \Xi_1 &= -\ii\hat\mu \otimes \hat\phi(f)\,,\qquad 
        \Xi_2 = -(\openone\otimes\openone)\Delta \,,\\
        \Xi_{j} &= 0\qquad \forall j\geq 3\,.
    \end{aligned}
\end{equation}
Here we defined $\Delta$ to be
\begin{subequations}
\begin{align}
    \Delta &= \frac{1}{2}\int\dd t\,\dd t'\Theta(t-t') \ii \Delta(t,t')\,,\\
    \Delta(t,t') &\coloneqq \int_{\Sigma_t}\!\!\!\dd^n\bx\int_{\Sigma_{t'}}\!\!\!\dd^n\bx' f(t,\bx)E(\sx,\sx')f(t',\bx')
\end{align}
\label{eq: delta-integral}
\end{subequations}
for some choice of Cauchy slice $\Sigma_t$ associated to some global time parameter $t$. Consequently, the joint unitary of the detector-field system reduces to 
\begin{align}
    \hat{U} = e^{-\ii\Delta}e^{-\ii\hat{\mu}\otimes\hat\phi(f)}\,,
\end{align}
Note that the extra phase $e^{-\ii\Delta}$ is a global phase that does not matter to the single-detector dynamics.

\subsection{Pure dephasing model}

The pure dephasing model is defined by some nonzero free Hamiltonian $\mathfrak{h}_0$ with nonzero $\Omega$, but instead the interaction Hamiltonian density chosen such that it commutes with the free Hamiltonian. That is, if we consider $\mathfrak{h}_0 \propto (\hat{\sigma}^z+\openone)$ as given in Eq.~\eqref{eq: free-hamiltonian}, then the interaction Hamiltonian is prescribed to be
\begin{align}
    \hat{h}_I(\sx) &= f(\sx)\hat{\sigma}^z \otimes \hat\phi(\sx)\,.
\end{align}
Since by construction we have $[\hat{\mathfrak{h}}_0,\hat{h}_I(\sx)]=0$, we have removed the time-dependence on the monopole operator since $\hat{\sigma}^z(\tau) = \hat{\sigma}^z$ for the above free Hamiltonian. Using the Magnus expansion, it is clear that the pure dephasing model has unitary time evolution given by
\begin{align}
    \hat{U} = e^{-\ii\Delta}e^{-\ii{\hat{\sigma}^z}\otimes\hat\phi(f)}\,.
\end{align}
where $\Delta$ is defined as per Eq.~\eqref{eq: delta-integral}.

\subsection{Simple-generated unitaries and comparison of each detector model}

Here we make brief comments on the comparison of the different regimes. First of all, since the global phases in the gapless and dephasing models are physically irrelevant (they cancel when we compute the time evolution in the density matrix formalism), all the unitaries are essentially of the form
\begin{align}
    \hat{U} \sim \exp\left[-\ii (\hat{O}\otimes\hat{R})\right]
    \label{eq: unitary-simple-generated}
\end{align}
where $\hat{O}$ acts on the detector's Hilbert space $\mathcal{H}_\textsc{d}\cong\C^2$ and $\hat{R}$ acts on the field's Hilbert space $\mathcal{H}_\phi$ (in a particular GNS representation). Given a set of linearly independent operators $\{\hat{O}_j\}$ acting on $\mathcal{H}_\textsc{d}$ and $\{\hat{R}_j\}$ acting on $\mathcal{H}_\phi$, the joint operator
\begin{align}
    \hat{\mathsf{H}} \coloneqq \sum_{j=1}^r c_j\hat{O}_j\otimes\hat{R}_j \qquad c_j\neq 0
\end{align}
is said to be of Schmidt rank $r$. The unitary of the form \eqref{eq: unitary-simple-generated} we considered are all generated by an operator with Schmidt rank $r=1$ --- they are sometimes called \textit{simple-generated unitaries} \cite{Simidzija2018nogo}. Simple-generated unitaries have been used extensively in the context of relativistic quantum communication channels \cite{tjoa2022channel,Landulfo2021cost}.

Clearly, the delta-coupling regime differs from the gapless and the pure dephasing regimes in that by construction the delta-coupling model restricts the kind of spacetime smearing allowed. Since in the rest frame of the qubit the interaction is effectively at a single instant in time, the delta-coupling regime is unable to capture non-perturbatively the dynamics of long-time processes such as thermalization and relaxation. This model is also in a way quite insensitive to the detector's trajectories since one needs to interact for sufficiently long times to obtain information about the trajectories. It is however possible to consider delta-coupling regime where the detector couples to the field $N$ times along its trajectory: such $N$-delta interactions will produce dynamics that are equivalent to collisional models in the open quantum systems literature \cite{Pipo2020spinboson}.

It is also clear that the pure dephasing model is very similar to the gapless model (as recognized in \cite{Landulfo2021cost}) because they both share the property that $[\hat{\mathfrak{h}}_0,\hat{h}_I(\sx)]=0$. However, they achieve this through different means and with different physical intuition in mind. In the gapless model, the intuition is that of a qubit detector whose gap is much smaller than all the frequency scales of the problem, hence its internal dynamics is much slower than the rest. Therefore, \textit{all} of the qubit observables essentially do not evolve in time in the interaction picture and we are free, in this context, to take the monopole operator to be any Hermitian operator $\hat{A}^\dagger=\hat{A}$. Note that because of effectively degenerate energy levels, a gapless detector model has no well-defined notion of \textit{thermal states}, since
\begin{align}
    \hat{\rho}_\textsc{d}(\beta) \coloneqq \frac{e^{-\beta\mathfrak{h}_0}}{\tr e^{-\beta\mathfrak{h}_0}} \equiv \frac{\openone}{2}\,.
\end{align}
In effect, the gapless detector can be interpreted, in a way, as going into the high-temperature regime $\beta\Omega\ll 1$. That said, if we also assume that the zero-gap Hamiltonian is an approximation of small energy gap, then despite the degenerate energy levels one would like to still think of the energy eigenstates as given by the eigenstates of say $\hat{\sigma}^z$. 

In contrast, since the pure-dephasing model has nonzero energy gap, thus it allows thermal states to be defined for the qubit detector. Unlike the gapless model, the pure dephasing model does not imply that all observables do not evolve, but rather one chooses the coupling to the field appropriately to make the interaction Hamiltonian commute with the free Hamiltonian of the qubit detector. The resulting dynamics due to the interaction with the field is \textit{functionally} similar to the gapless model, but the model speaks about different physical setups. Indeed, as far as the dephasing behaviour is concerned, there is no distinction between the gapless and pure dephasing models, thus in what follows we will always favor the pure dephasing model and all results will carry through for the gapless model. 

Since the global phase in the gapless and pure dephasing models will not matter, it is now convenient to unify all three regimes since we always have a simple-generated unitary with Schmidt rank $r=1$. Let us define a universal simple-generated unitary for all three models, given by
\begin{align}
    \hat{U} &\coloneqq \exp[-\ii \hat{O}\otimes\hat{\phi}(f)]\,, 
\end{align}
where it is understood that $\hat{O} = \hat{\sigma}^x(\tau_0),\hat{\mu},\hat{\sigma}^z$ for the monopole operator of the delta-coupled, gapless, and pure dephasing models respectively\footnote{Note that for gapless model the monopole $\hat{\mu}$ could be \textit{any} Hermitian field observables.}. The smeared field operator $\hat{\phi}(f)$ corresponds to the choice of spacetime smearing $f$ chosen appropriately (given by \eqref{eq: delta-smearing} for delta coupling regime, and any $f\in\CS$ for gapless and dephasing models). The universal form of the unitary will enable us to deal with all three models simultaneously when we construct the the relativistic quantum channels for both the qubit and the field respectively in the next section.

\section{Quantum channels induced by simple-generated unitary evolution}
\label{sec: channels}

In this section we construct the qubit channel $\Phi$ induced by the non-perturbative interactions with the field. We will also construct the so-called the \textit{complementary channel} $\Phi^c$ associated to $\Phi$ that we will define below. The goal of our calculations is to extend our results to include field states that can be obtained from a quasifree state by Gaussian operations such as displacement and squeezing. 


Let $\mathscr{D}(\mathcal{H})$ be the space of density operators acting on some Hilbert space $\mathcal{H}$, and let $\mathcal{H}_\textsc{d}$ and $\mathcal{H}_\phi$ be the Hilbert spaces of the qubit detector and the field respectively. In what follows we assume that the detector and the field is prepared initially in some uncorrelated state
\begin{align}
    \hat{\rho}_{\textsc{d}\phi}^0 = \rao\otimes\rof\,. 
\end{align}
The joint unitary dynamics given by the simple-generated Schmidt rank-1 unitary $\hat U$ \eqref{eq: unitary-simple-generated} gives rise to a quantum channel $\Phi:\mathscr{D}(\mathcal{H}_\textsc{d})\to \mathscr{D}(\mathcal{H}_\textsc{d})$, which reads
\begin{subequations}    
\begin{align}
    \Phi(\rao) &= \tr_\phi(\hat{U}(\rao\otimes\rof)\hat{U}^\dagger)\,.
    \label{eq: detector-channel}
\end{align}
The corresponding complementary channel is defined to be $\Phi^c:\mathscr{D}(\mathcal{H}_\textsc{d})\to\mathscr{D}(\mathcal{H}_\phi)$, which reads
\begin{align}
    \Phi^c(\rao) &= \tr_\textsc{d}(\hat{U}(\rao\otimes\rof)\hat{U}^\dagger)\,.
    \label{eq: complementary-channel}
\end{align}
Related to these two channels, we can define another quantum channel acting on the field state $\tilde{\Phi}:\mathscr{D}(\mathcal{H}_\phi)\to\mathscr{D}(\mathcal{H}_\phi)$, given by
\begin{align}
    \tilde{\Phi}(\rof) &= \tr_\textsc{d}(\hat{U}(\rao\otimes\rof)\hat{U}^\dagger)\,.
    \label{eq: environment-channel}
\end{align}
\end{subequations}
Note that Eq.~\eqref{eq: complementary-channel} is distinct from Eq.~\eqref{eq: environment-channel} in that the complementary channel $\Phi^c$ is defined for a \textit{fixed} initial field state $\rof$, while the channel $\tilde\Phi$ is defined for a \textit{fixed} qubit initial state $\rod$. Furthermore, it is now clear from these definitions that the global phase factor $e^{-\ii\Delta}$ that appear in the gapless and pure dephasing models drop out of the calculations, hence the universal simple-generated unitary \eqref{eq: unitary-simple-generated} suffices to specify these channels.

In order to obtain closed-form expressions for these channels, it is convenient to rewrite the unitary as a finite sum of bounded operators
\begin{align}
    \hat{U} =\openone\otimes \cos \hat\phi(f)-\ii\hat{O}\otimes \sin\hat \phi(f)\,.
    \label{eq: unitary-universal}
\end{align}
By writing $C_{f}=\cos\hat\phi({f})$ and $S_{f} = \sin\hat\phi({f})$, the joint state after the unitary evolution reads
\begin{align}
    \hat{\rho}_{\textsc{d}\phi} &= \hat{U}\hat{\rho}^0_{\textsc{d}\phi} \hat{U}^\dagger \notag\\
    &= \rod \otimes C_{f}\rof C_{f} +\hat{O} \rod \hat{O}  \otimes S_{f}\rof S_{f} \notag\\
    &\hspace{0.4cm}- \ii \hat{O} \rod\otimes S_{f} \rof C_{f} + \ii \rod \hat{O}  \otimes C_{f}\rof S_{f}\,.
    \label{eq: algebraic-state-total}
\end{align}
From this expression we can obtain the three quantum channels and the resulting actions on the field and detector states as given in Eqs.~\eqref{eq: detector-channel}=\eqref{eq: environment-channel}.

\subsection{The qubit channel $\Phi$}
First, let us calculate the final state of the detector $\hat{\rho}_\textsc{d}$ after the interaction. This is given by $\hat{\rho}_\textsc{d} = \Phi(\rod)$, which reads
\begin{align}
    \Phi(\rod)
    &=  \omega(C_{f}^2) \rod + \omega(S_{f}^2) \hat{O}\rod\hat{O}  + \ii\omega(S_fC_f)[\rod,\hat{O}]\,,
\end{align}
where we have used the fact that we can write\footnote{This follows from the fact that algebraic states assign expectation value of observables \cite{fewster2019algebraic}. Alternatively, we can think of the RHS as being evaluated in the GNS representation associated to the algebra of observables $\A(\M)$ and the state $\omega$.}
\begin{align}
    \omega(A) \equiv \tr_\phi(\rof\hat A)\,.
    \label{eq: formal-field-trace}
\end{align}
The cyclic property of the trace makes sense since $C_f,S_f$ are bounded operators. For our purposes this expression is good enough, but from the perspective of quantum channel theory it pays to express the channel in its Kraus representation. The idea is to rewrite the unitary $\hat{U}$ in terms of eigenprojectors of $\hat{O}$: let $\hat{P}_\pm \coloneqq \frac{1}{2}(\openone\pm \hat{O})$ with eigenvalues $p_\pm = \pm1$, so that we have
\begin{align}
    \hat{U} &= \hat{P}_- \otimes e^{\ii\hat\phi(f)}+\hat{P}_+\otimes e^{-\ii\hat\phi(f)}\,.
\end{align}
Following the convention in \cite{Landulfo2016magnus1,tjoa2022channel}, let us define 
\begin{align}
    \nu_f \coloneqq \omega(e^{2\ii\hat\phi(f)})\in \mathbb{C}\,.
\end{align}
We can now rewrite the channel in its Kraus representation
\begin{align}
    \Phi(\rod) &= \sum_{j=0}^2 \mathcal{K}_j\rod \mathcal{K}_j^\dagger\,,
    \label{eq: bit-flip-delta-kraus}
\end{align}
where the Kraus operators are
\begin{equation}
    \begin{aligned}
    \mathcal{K}_0 &= \sqrt{\frac{1-|\nu_f|}{2}}\openone\,,\qquad \mathcal{K}_1 = \sqrt{\frac{1-|\nu_f|}{2}}\hat{O}
    \,,\\
    \mathcal{K}_2 &= \sqrt{\frac{|\nu_f|+\Re\,\nu_f}{2}}\openone - \ii \sqrt{\frac{|\nu_f|-\Re\,\nu_f}{2}}\hat{O}\,.
    \end{aligned}
\end{equation}
In order to make connections with the expressions involving $C_f$ and $S_f$, it is useful to note that
\begin{equation}
\begin{aligned}
    \omega(C_{f}) &= \frac{1}{2}\left(\omega(e^{\ii\hat{\phi(f)}})+\omega(e^{-\ii\hat{\phi(f)}})\right) = \Re\,\nu_{f/2}\,,\\
    \omega(S_{f}) &= \frac{1}{2\ii}\left(\omega(e^{\ii\hat{\phi(f)}})-\omega(e^{-\ii\hat{\phi(f)}})\right) = \Im\,\nu_{f/2}\,.
\end{aligned}
\end{equation}
We can then use `functional calculus'  in the sense that $C_f,S_f$ can be evaluated as if they are the usual trigonometric functions: for example, we have
\begin{equation}
    \begin{aligned}
        \omega(C_f^2) &= \frac{1}{2}(1+C_{2f}) = \frac{1+\Re\,\nu_f}{2}\,,\\
        \omega(S_f^2) &= \frac{1}{2}(1-C_{2f}) = \frac{1-\Re\,\nu_f}{2}\,,\\
        \omega(C_fS_f) &= \frac{1}{2}S_{2f} = \frac{\Im\,\nu_f}{2}\,.
    \end{aligned}
    \label{eq: nuf}
\end{equation}
We can now see that the restriction of $\omega$ to the class of quasifree states (\textit{c.f.} Section~\ref{sec: AQFT}) gives us $\nu_f=e^{-2\mathsf{W}(f,f)}\in (0,1]$, and in particular $\nu_f$ is real-valued. In this work part of our goal is to avoid this restriction and extend the standard calculations involving vacuum state of the field (which is quasifree) to much more general class of Gaussian states that may not be quasifree, such as the (squeezed) coherent state.

Using these expressions, the action of the quantum channel now reads
\begin{align}
    \Phi(\rod)
    &=  \frac{1+\Re\,\nu_f}{2} \rod + \frac{1-\Re\,\nu_f}{2} \hat{O}\rod\hat{O}  \notag\\
    &\hspace{2.35cm} 
    + \ii\frac{\Im\,\nu_f}{2}[\rod,\hat{O}]\,.
    \label{eq: bit-flip-universal}
\end{align}
In the quasifree case we can give a very clean interpretation of this qubit channel. Consider the case where $\hat{O} = \hat{\sigma}^x$ (in the delta-coupling case we can always adjust $\Omega\tau_0$ to obtain this). If the field state is quasifree, then $\nu_f\in (0,1]$ and the final state (in the interaction picture) simplifies to 
\begin{align}
     \Phi(\rod)\Bigr|_{\text{qf}}
    &=  \frac{1+e^{-2\mathsf{W}(f,f)}}{2} \rod + \frac{1-e^{-2\mathsf{W}(f,f)}}{2} \hat{\sigma}^x\rod\hat{\sigma}^x\,,
    \label{eq: quasifree-qubit}
\end{align}
which is nothing but the \textit{bit-flip channel}. Furthermore, this tells us that when the fluctuations of the field is large, i.e., $\mathsf{W}(f,f)\gg 1$, then the channel is so noisy that in effect we lose all information about the field: the channel reduces to uniform random bit-flip in the limit of large fluctuations $\mathsf{W}(f,f)\to \infty$. If we now consider $\hat{O}=\hat{\sigma}^z$ as in the pure dephasing (or gapless) model, the channel reduces to the \textit{phase-flip channel}. Furthermore, in the limit of large fluctuations the channel becomes a \textit{completely dephasing channel}, thus we also lose all information about the field. 

It is interesting to note that the regime of large fluctuations $\mathsf{W}(f,f)\gg 1$ can be attained in several ways. For example, we can consider very sharply localized $f$, since in the limit $f\to \delta^{n+1}(\sx)$ the Wightman two-point function is ultraviolet (UV)-divergent. Alternatively, we can increase the coupling strength $\lambda$ of the detector-field interaction. Since $\mathsf{W}(f,f)$ scales with $\lambda^2$, the value of $\nu_f = e^{-2\mathsf{W}(f,f)}$ decays exponentially with $\lambda^2$. Consequently, the strong coupling quickly also erases all information about the field from the detector's (final) state. This is consistent with the physical intuition that strong coupling regime is equivalent to the UV (high-energy) regime. In contrast, in the limit of small fluctuations (e.g., using careful adiabatically switched detectors, or by going to the very weak coupling $\lambda\to 0$), we have $\mathsf{W}(f,f) \to 0$ and hence the resulting channel approaches a \textit{noiseless channel} $\Phi(\rod) \to \rod$.

\subsection{Output of the complementary channel $\Phi^c$ and the field channel $\tilde{\Phi}$}
\label{sec: complementary}

Arguably, the more interesting channels are the complementary channel $\Phi^c$ and also the field channel $\tilde{\Phi}$ since they are much less studied in the RQI literature. This is because the Hilbert space of the field is infinite-dimensional, and in free space there are uncountably many degrees of freedom (unlike quantum harmonic oscillators). This makes concrete calculations much more difficult. There are some known results: for example, the complementary channel $\Phi^c$ is an entanglement-breaking channel \cite{Simidzija2018nogo}, which is the reason why the quantum communication channels based on simple-generated interactions considered in \cite{Landulfo2016magnus1,tjoa2022channel} cannot transmit quantum information (the quantum channel capacity is zero).

At this stage, it is quite difficult to study the channels directly but we can learn something about these channels by studying their output field states. That is,  after the interaction the field state is given by
\begin{align}
    \hat{\rho}_\phi &= \Phi^c(\rod) = \tilde{\Phi}(\rof)\,.
    \label{eq: field-output}
\end{align}
It is worth stressing that the channels themselves are not equivalent: for example, $\tilde{\Phi}$ will not be entanglement-breaking while $\Phi^c$ is entanglement-breaking\footnote{This has to do with the fact that an entanglement-breaking channel $\mathcal{E}$ can be recast in the form $\mathcal{E}(\rho) = \sum_j\tr(E_j\rho)\hat{\sigma}_j$ where $\hat{E}_j$ are POVM elements and $\hat{\sigma}_j$ are density operators. Suitably generalized to the infinite-dimensional case, we see that $\tilde{\Phi}$ cannot be put in this form while $\Phi^c$ readily does.}. For our purposes, however, we would like to reframe Eq.~\eqref{eq: field-output} in terms of the algebraic state $\omega$ so that we can express our result in a representation-independent manner.

The idea goes as follows. First, from Eq.~\eqref{eq: algebraic-state-total} we  write the output state $\hat{\rho}_\phi$ as the action of the channel $\Tilde{\Phi}(\rof)$, which reads
\begin{align}
    \hat{\rho}_\phi \equiv \tilde{\Phi}(\rof) 
    &= C_f\rof C_f+S_f\rof S_f
    \notag\\
    &\hspace{1cm} 
    + \ii \braket{\hat{O}}(C_f\rof S_f-S_f\rof C_f)\,, 
    \label{eq: field-state-general}
\end{align}
where $ \braket{\hat{O}} = {\tr}(\hat{O}\rod)$. Using Eq.~\eqref{eq: formal-field-trace}, we can think of the action of $\tilde{\Phi}$ as mapping from the initial algebraic state $\omega$ to a new algebraic state $\omega'$, such that for any $A\in \W(\M)$ we have
\begin{align}
    \omega'(A)&=
    \omega(C_fAC_f)+\omega(S_fAS_f) \notag\\
    &\hspace{1.5cm} + \ii\braket{\hat{O}}(\omega(S_fAC_f)-\omega(C_f A S_f))\,.
    \label{eq: algebraic-state-new-0}
\end{align}
Each of these terms can be evaluated purely using the Weyl relations of $\W(\M)$.

For completeness, let us show that the state $\omega'$ is a {mixed} state in accordance to the algebraic definition.
One simple way to see this is to consider the special case when $\braket{\hat{O}}=0$ (by choosing a suitable state of the detector). For $\braket{\hat{O}}=0$ we get
\begin{align}
    \omega'(A) &=  \omega(C_fAC_f)+\omega(S_fAS_f)\,.
    \label{eq: quasifree-final}
\end{align}
Now we define two algebraic states $\omega_1$ and $\omega_2$:
\begin{align}
    \omega_1(A) &= \frac{\omega(C_fAC_f)}{\omega(C_f^2)}\,,\quad \omega_2(A) = \frac{\omega(S_fAS_f)}{\omega(S_f^2)}\,,
\end{align}
we can rewrite $\omega'$ as a convex combination
\begin{align}
    \omega' &= \omega(C_f^2)\omega_1+\omega(S_f^2)\omega_2\,,
    \label{eq: algebraic-state-mixed}
\end{align}
with $\omega(C_f^2)+\omega(S_f^2) = 1$. Since a state is pure if and only if it cannot be written as a strict convex combination of two algebraic states \cite{Khavkhine2015AQFT}, it follows that $\omega'$ is mixed, as we expect. Furthermore, this means that the detector and the field are necessarily entangled after the interaction. 



\section{Gaussian operations as adjoint channels on observables}
\label{sec: gaussian-operations}

In this section we will exploit the concept of adjoint channel, which is essentially the formulation of quantum channels in the ``Heisenberg picture'', to encode Gaussian operations such as displacement and squeezing. This will enable us to reformulate the UDW-type interactions with the field in a Gaussian but non-quasifree state in terms of known results using quasifree states. In effect, this allows us to provide a ``configuration space'' reinterpretation of the displacement and squeezing operations in terms of the detector observables.

\subsection{Displacement and squeezing operations}

In UDW settings, one of the most common choices for the field's initial state is the vacuum state, which we denote here by $\omega_0$. This is only one of the many classes of Gaussian states (fully characterized by one-point and two-point functions) in quantum field theory. There are at least three types of non-vacuum states that are of great interest: thermal states, squeezed vacuum states, and coherent states. Thermal and squeezed vacuum are quasifree states (vanishing one-point functions), while coherent states are Gaussian states that are not quasifree. We can construct more Gaussian states by series of coherent displacement and squeezing operations on any Gaussian states. Below we will show that we can rephrase coherent and squeezing operators as elements of the Weyl algebra\footnote{We are not the first to regard coherent states and squeezed states in QFT this way (see e.g. \cite{lashkari2021modular,Hollands2019news,Casini2019coherent}), though usually this is framed ``backwards'': they \textit{define} coherent and squeezed states directly via $e^{\ii\hat\phi(g)}$ and $e^{\ii\hat\phi(g)^2}$ respectively instead of a more optically-motivated definition via ladder operators $\hat a_\bk^{\phantom{\dagger}},\hat a_\bk^\dagger$. }, which also provides us with straightforward generalization of quasifree calculations with minimal effort.

A generic coherent state is given by the displacement operator acting on the vacuum state:
\begin{align}
    \ket{\alpha}\coloneqq \hat{D}(\alpha)\ket{0}\,,
\end{align}
where $\alpha$ is the coherent amplitude (which is typically multimode, see \cite{simidzija2018harvesting}) and $\hat{D}(\alpha)$ is the displacement operator. In the Fock space representation induced by the GNS theorem, it reads
\begin{align}
    \hat{D}(\alpha) &= e^{\int\dd^n\bk (\alpha(\bk)\hat a_\bk - \alpha(\bk)^*\hat a_\bk^\dagger)}\,.
\end{align}
Observe that we can take some $\tilde{\alpha} \in \CS$ and define
\begin{align}
    \alpha(\bk)\coloneqq \ii \int \dd V\,\tilde{\alpha}(\sx)u_\bk(\sx)\,.
    \label{eq: coherent-amplitude}
\end{align}
By construction $\alpha\in L^2(\R^n)$, i.e.,  $\int\dd^n\bk\,|\alpha(\bk)|^2<\infty$ since $\tilde{\alpha}(\sx)$ is compactly supported smooth function, although $\alpha(\bk)$ will not in general be compactly supported. Eq.~\eqref{eq: coherent-amplitude} implies that we can view
\begin{align}
    \hat{D}(\alpha) &= e^{\ii\hat\phi(\tilde{\alpha})}\in \W(\M)
    \label{eq: displacement-weyl}
\end{align}
for some $\tilde{\alpha}\in \CS$. Note that we have used the notation $\tilde{\alpha}$ to make suggestive analogy to Fourier transform (which is indeed the Fourier transform when $\M$ is Minkowski space up to a prefactor $\ii$).  

With similar approach, we can also define squeezed vacuum state by the action \cite{simidzija2018harvesting},
\begin{align}
    \ket{\hat{S}(\zeta)}\coloneqq \hat{S}(\zeta)\ket{0}\,,
\end{align}
where in the Fock representation it is given by
\begin{align}
    \hat{S}(\zeta) &= e^{\frac{1}{2}\int\dd^n\bk\,\dd^n\bk' (\zeta(\bk,\bk')\hat a_\bk \hat a_{\bk'}- \text{H.c.})}\,.
\end{align}
Now let $\tilde{\zeta}\in \CS$, using suggestive notation as before, and define
\begin{align}
    \zeta(\bk,\bk')\coloneqq 2\ii\int\dd V\,\dd V' \tilde{\zeta}(\sx)\tilde{\zeta}(\sx')u_\bk(\sx)u_{\bk'}(\sx')\,.
\end{align}
We assume in this case that the squeezing amplitudes $\zeta(\bk,\bk')$ can be written in this way\footnote{\textit{A priori} this is not covering all possible squeezing operations, since it includes possibly momentum-entangling squeezing where $\zeta(\bk,\bk')\neq \zeta_1(\bk)\zeta_2(\bk')$ for some functions $\zeta_j$. We restrict our attention to this subclass for simplicity.}, hence as before we can view the squeezing operator as an element of the Weyl algebra, namely
\begin{align}
    \hat{S}(\zeta) = e^{\ii\hat\phi(\tilde\zeta)^2}\in\W(\M)\,.
\end{align}
In other words, it is the exponentiation of bi-local smeared operator $\hat\phi(\tilde\zeta)^2\in\A(\M)$. This covers a large class of squeezing operations we are interested in. 

By interpreting these operations as Weyl elements, we can define an adjoint channel associated to the displacement and squeezing operations, essentially moving into the ``Heisenberg picture'' form of these operations. Suppose we consider a coherent state $\ket\alpha$. At the level of the GNS representation of the vacuum state $\omega_0$, we have the coherent state density matrix $\hat{\rho}^0_{\phi,\alpha}\coloneqq \hat{D}(\alpha)\rof \hat{D}(\alpha)^\dagger$ where $\rof=\ketbra{0}{0}$. More generally, we can treat displacement operation as a unitary channel $\mathcal{U}_\alpha:\mathscr{D}(\mathcal{H}_\phi)\to \mathscr{D}(\mathcal{H}_\phi)$ given by
\begin{align}
    \mathcal{U}_\alpha(\rof) = \hat{D}(\alpha)\rof \hat{D}(\alpha)^\dagger\,.
\end{align}
It follows then
\begin{align}
    \omega_\alpha(A) &=  \tr(\hat{\rho}^0_{
    \phi,\alpha}\hat A) = \tr(\rof \hat{D}(\alpha)^\dagger \hat A \hat{D}(\alpha)) \notag\\
    &\equiv  \omega_0(\mathcal{U}^\dagger_\alpha(A))\,,
    \label{eq: coherent-algebraic-state}
\end{align}
where $\mathcal{U}^\dagger_\alpha(\cdot) \coloneqq \hat{D}(\alpha)^\dagger (\cdot ) \hat{D}(\alpha)$ is the adjoint channel of $\mathcal{U}_\alpha$  and $\omega_0$ is the vacuum state. For squeezed vacuum state we have the unitary squeezing channel
\begin{align}
    \mathcal{V}_\zeta(\rof) = \hat{S}(\zeta)\rof \hat{S}(\zeta)^\dagger\,.
\end{align}
The squeezed vacuum state is $\hat{\rho}^0_{\phi,\zeta}\coloneqq \hat{S}(\zeta)\rof \hat{S}(\zeta)^\dagger$ with $\rof=\ketbra{0}{0}$ and  it follows that 
\begin{align}
    \omega_\zeta(A) &=  \tr(\rof \hat{S}(\zeta)^\dagger \hat A \hat{S}(\zeta)) = \omega_0(\mathcal{V}^\dagger_\zeta (A))\,,
    \label{eq: squeezed-algebraic-state}
\end{align}
where $\mathcal{V}^\dagger_\zeta(\cdot )  \coloneqq \hat{S}(\zeta)^\dagger(\cdot ) \hat{S}(\zeta)$ is the corresponding adjoint channel. These conversions seem to be a very trivial move, however we will see in the next subsection that it is precisely this step that allows us to generalize various calculations to coherent and squeezed states despite not being quasifree. This is because we do know how to take expectation values with respect to $\omega_0$ based on the definition of the quasifree state \eqref{eq: quasifree-definition}. 

Since $\hat{D}(\alpha)$ and $\hat{S}(\zeta)$ are both unitary elements of the Weyl algebra, it is straightforward to calculate how the local noise of the field given by the symmetrically-smeared Wightman function changes with these operations. For coherent states, we have
\begin{align}
    \mathsf{W}_\alpha(f,f) &= \omega_\alpha(\hat\phi(f)\hat\phi(f)) \notag\\
    &= \omega_0(\hat{D}(\alpha)^\dagger \hat\phi(f)\hat\phi(f)\hat{D}(\alpha))\,.
\end{align}
However, using Eq.~\eqref{eq: displacement-weyl}, the Baker-Campbell-Hausdorff (BCH) formula and the CCR we have
\begin{align}
    \hat{D}(\alpha)^\dagger \hat\phi(f)\hat{D}(\alpha) &= \hat\phi(f) -\ii[\hat\phi(\tilde\alpha),\hat\phi(f)]\notag\\
    &= \hat\phi(f) + E(\tilde\alpha,f)\openone\,. 
    \label{eq: displacement-field}
\end{align}
Therefore, we get
\begin{align}
    \mathsf{W}_\alpha(f,f) &= \mathsf{W}_0(f,f) + E(\tilde\alpha,f)^2 \geq \mathsf{W}_0(f,f)\,.
\end{align}
We immediately get the result that the coherent state has larger noise contribution than the vacuum state by an amount that depends on the causal propagator between $\tilde\alpha$ and $f$, and furthermore this is only nonzero if $\Tilde{\alpha}$ and $f$ have supports that are causally connected by the field. This can be understood as follows: since coherent states are not invariant under the full spacetime isometry group (in flat space it is the Poincar\'e group), the coherent amplitude's ``Fourier transform'' must be localized somewhere in spacetime with support given by that of $\tilde\alpha$, and it is here that the coherent excitations add to the field fluctuations.

Similarly, for squeezed state we get 
\begin{align}
    \hat{S}(\zeta)^\dagger \hat\phi(f) \hat{S}(\zeta) &= \hat\phi(f) -\ii[\hat\phi(\tilde\zeta)^2,\hat\phi(f)]\notag\\
    &= \hat\phi(f) + 2 E(\tilde\zeta,f)\hat\phi(\tilde{\zeta})\notag\\
    &\equiv \hat\phi(h_\zeta)\,.
    \label{eq: squeezing-field}
\end{align}
This has the nice interpretation that the squeezing operation ``squeezes the smearing profile'' $f$ into $h_\zeta\coloneqq f+2E(\tilde{\zeta},f)$. Therefore, we get
\begin{align}
    \mathsf{W}_\zeta(f,f) &= \mathsf{W}_0(f,f) + 2E(\tilde\zeta,f)\omega_0(\{\hat\phi(\tilde\zeta),\hat\phi(f)\})\notag\\
    &\hspace{1cm}+4E(\tilde{\zeta},f)^2\mathsf{W}_0(\tilde{\zeta},\tilde{\zeta})\,.
    \label{eq: squeezed-wightman}
\end{align}
Unlike the case for coherent state, it is no longer the case that any choice of squeezing amplitude leads to larger noise than the vacuum since the second term is not positive semidefinite. What remains true, however, is that again the impact of squeezing on the field fluctuations is not uniform in spacetime, since it is controlled by the ``Fourier transform'' $\tilde\zeta$. Observe that in flat space where $\tilde\zeta$ is indeed the Fourier transform of $\zeta$, we can choose $L^2$-integrable function $\zeta$ with $\tilde\zeta$ compactly supported, so that the squeezing only impacts the region $\supp(\tilde\zeta)$ causally connected to interaction region $f$. A version of this spatial dependence of squeezing on detector dynamics in flat spacetime was analyzed in \cite{simidzija2018harvesting}.

This calculation generalizes to multiple displacement and squeezing without having to solve any momentum integrals: for example, if we consider 
\begin{align}
    \ket{\beta+\alpha}\coloneqq \hat{D}(\beta)\hat{D}(\alpha)\ket{0}\,,
\end{align}
we see that Eq.~\eqref{eq: displacement-field}
\begin{align}
    \hat{D}(\alpha)^\dagger& \hat{D}(\beta)^\dagger\hat\phi(f) \hat{D}(\beta)\hat{D}(\alpha)\notag\\
    &= \hat{D}(\alpha)^\dagger \hat\phi(f) \hat{D}(\alpha) + E(\tilde\beta,f)\openone\notag\\
    &= \hat\phi(f)+E(\tilde\alpha+\tilde\beta,f)\openone\notag\\
    &= \hat{D}(\alpha+\beta)^\dagger\hat\phi(f)\hat{D}(\alpha+\beta)\,.
    \label{eq: displacement-sequence}
\end{align}
In the second equality we have used the linearity of the causal propagator. Thus we see that arbitrary sequence of displacement operators does not pose any extra effort. Note that a sequence of squeezing operations is straightforward because the state remains quasifree: we have
\begin{subequations}
\begin{align}
    \mathcal{V}_{\eta}^\dagger\circ\mathcal{V}_\zeta^\dagger(\hat\phi(f)) &= \mathcal{V}_\eta(\hat\phi(h_\zeta)) = \hat\phi(h_{\eta\cdot\zeta})\,,
    \\
    h_{\eta\cdot\zeta} &\coloneqq h_\zeta + 2E(\tilde\eta,h_\zeta)\,,
\end{align}
\end{subequations}
with $h_\zeta = f+2E(\tilde\zeta,f)$.

As a more non-trivial example, consider a squeezed coherent state $\ket{\zeta;\alpha}\coloneqq \hat{S}(\zeta)\hat{D}(\alpha)\ket{0}$. Using Eq.~\eqref{eq: displacement-field} and Eq.~\eqref{eq: squeezing-field} we get
\begin{align}
    &\hat{D}(\alpha)^\dagger \hat{S}(\zeta)^\dagger\hat\phi(f) \hat{S}(\zeta)\hat{D}(\alpha)\notag\\
    &= \hat{D}(\alpha)^\dagger \hat\phi(f) \hat{D}(\alpha) + \hat{D}(\alpha)^\dagger (2E(\tilde\zeta,f)\hat\phi(\tilde\zeta))\hat{D}(\alpha)\notag\\
    &= \hat\phi(f)+E(\tilde\alpha,f)\openone+2E(\tilde\zeta,f)\hat\phi(\tilde\zeta) + 2E(\tilde\zeta,f)E(\tilde\alpha,f)\openone\notag\\
    &= \hat\phi(h_\zeta) + E(\tilde\alpha,h_\zeta)\openone\,,
    \label{eq: deformed-coherent}
\end{align}
where $h_\zeta\coloneqq f+2E(\tilde\zeta,f)\tilde\zeta$. Eq.~\eqref{eq: deformed-coherent} suggests that the action of adjoint coherent and squeezing channels on $\hat\phi(f)$ is equivalent to displacement operator acting on ``deformed smearing'' $\hat\phi(h_\zeta)$, and hence, the two unitary adjoint channels commute up to a phase:
\begin{align}
    (\mathcal{U}^\dagger_\alpha\circ\mathcal{V}^\dagger_\zeta-\mathcal{V}^\dagger_\zeta\circ\mathcal{U}^\dagger_\alpha)(\hat \phi(f)) &= E(\tilde\zeta,f)E(\tilde\alpha, \tilde\zeta)\openone\,.
\end{align}
Compared to the usual momentum-space calculations involving ladder operators, this computation is manifestly simpler. Furthermore, by definition of adjoint we also get the same result for the state, i.e., displacement and squeezing acting on the state commutes up to a phase. However, there is something less obvious that we can glean from this: the fact that the phase depends on $E(\tilde\alpha,\tilde{\zeta})$ shows that squeezing and displacement commutes iff the ``Fourier transforms'' of the displacement and squeezing amplitude $\tilde\alpha,\tilde\zeta$ are \textit{causally disconnected} with respect to the field. Indeed, if two regions are spacelike-separated and one observer adds coherent excitations to one region and the other performs local squeezing, then their operations should not influence one another. 


\subsection{Qubit channel revisited: coherent and squeezed states}

The expressions we just obtained give us a very straightforward generalization  of the qubit channel in Section~\ref{sec: channels} to two important classes of non-quasifree Gaussian states, namely for algebraic states associated to coherent and squeezed states. This works because we can treat the coherent and squeezing as adjoint channel acting on the observable elements, and then take expectation values with respect to a reference quasifree state (such as the vacuum). Since expectation values associated to quasifree states are given directly in terms of the Wightman two-point functions, the change into Heisenberg picture gives us a way to algebraically generalize the qubit channel calculations into more general non-quasifree (but Gaussian) settings. 

More concretely, for coherent state with coherent amplitude $\alpha$, we have
\begin{align}
    \Phi_\alpha(\rod) &= \omega_\alpha(C_{f}^2) \rod + \omega_\alpha(S_{f}^2) \hat{O}\rod\hat{O}  \notag\\
    &\hspace{1.9cm}+ \ii[\rod,\hat{O}]\omega_\alpha(S_fC_f)\,.
    \label{eq: coherent-qubit-channel}
\end{align}
Using the adjoint channel \eqref{eq: coherent-algebraic-state}, we have
\begin{subequations}
\begin{align}
    \omega_\alpha(C_f^2) &= \omega_0(C_{\tilde{\alpha}}C_f^2C_{\tilde{\alpha}}) + \omega_0(S_{\tilde{\alpha}}C_f^2S_{\tilde{\alpha}}) \notag\\
    &= \frac{1+\nu_f\cos(2E(\tilde\alpha,f))}{2}\,,\\
    \omega_\alpha(S_f^2) &= \omega_0(C_{\tilde{\alpha}}S_f^2C_{\tilde{\alpha}}) + \omega_0(S_{\tilde{\alpha}}S_f^2S_{\tilde{\alpha}}) \notag\\
    &= \frac{1-\nu_f\cos(2E(\tilde\alpha,f))}{2}\,,\\
    \omega_\alpha(S_fC_f) &= -\ii\omega_0(S_{\tilde\alpha}S_fC_fC_{\tilde\alpha})+\ii\omega_0(C_{\tilde\alpha}S_fC_fS_{\tilde\alpha})
    \notag\\
    &= \frac{1}{2}\nu_f
    \sin(2E(\tilde\alpha,f))\,.
    \label{eq: coherent-sine-cos}
\end{align}
\end{subequations}
where $\nu_f = e^{-2\mathsf{W}_0(f,f)}$.  As expected,  unlike the quasifree case $\omega_\alpha(S_fC_f)$ is no longer zero. These can be straightforwardly computed by direct computation using Weyl relations, or more neatly using ``trigonometric lemma'' in \cite[Lemma 1]{Tjoa2022fermi}, which we quote for convenience:
\begin{subequations}
    \begin{align}
        2C_iC_j &= C_{i+j}e^{-\ii E_{ij}/2} + C_{i-j}e^{\ii E_{ij}/2} \,,\\
        -2S_iS_j &= C_{i+j}e^{-\ii E_{ij}/2} - C_{i-j}e^{\ii E_{ij}/2} \,,\\
        2C_iS_j &= S_{i+j}e^{-\ii E_{ij}/2} - S_{i-j}e^{\ii E_{ij}/2} \,,\\
        2S_iC_j &= S_{i+j}e^{-\ii E_{ij}/2} + S_{i-j}e^{\ii E_{ij}/2} \,,
    \end{align}
\end{subequations}
where $C_{i\pm j}\equiv \cos\hat\phi(f_i\pm f_{j})$, $S_{i\pm j}\equiv \sin\hat\phi(f_i\pm f_{j})$ and $E_{ij}\coloneqq E(f_i,f_j)$ is the smeared causal propagator.

It is interesting to observe that the coefficients in Eq.~\eqref{eq: coherent-qubit-channel} can also be computed in the same way as what goes into the calculations involving two-qubit communication settings (see, e.g., Eq.~(44)-(49) of \cite{Tjoa2022fermi}). The resulting expression appears as some sort of modulation of $\nu_f$ that appears in Eq.~\eqref{eq: bit-flip-universal} by the sine and cosine of the causal propagator, and we reproduce the vacuum result (which is quasifree) when ${\alpha}\to 0$. This also shows that the coherent state of the field modify expectation values of the field state non-uniformly in spacetime, which is to be expected since physically meaningful coherent states must have excitations that are sufficiently localized in spacetime and its influence is propagated by the field via the propagator $E$. In particular, this leads to the nice interpretation that coherent states are localized in such a way that any observer/detector that are causally disconnected from $\tilde\alpha$ will not ``feel'' the coherent excitations and view the field as being essentially the vacuum state.

For squeezed states, we can perform analogous calculation but we will have to work out the coefficients more directly as follows. Using the adjoint channel \eqref{eq: squeezed-algebraic-state} and the BCH formula we have
\begin{align}
    \hat{S}(\zeta)^\dagger e^{\ii\phi(f)} \hat{S}(\zeta) &=e^{\ii(\hat\phi(f) -2E(f,\tilde\zeta)\hat\phi(\tilde\zeta))}\,.
\end{align}
Therefore, we have for instance
\begin{align}
    \omega_\zeta(C_f^2) 
    &= \frac{1}{2} + \frac{1}{2}\omega_\zeta(e^{2\ii\hat\phi(f)}+e^{-2\ii\hat\phi(f)})\notag\\
    &= \frac{1}{2}+\frac{1}{2}\bigr(\omega_0(e^{2\ii\hat\phi(h_+)})+\omega_0(e^{-2\ii\hat\phi(h_-)})\bigr)\notag\\
    &=\frac{1+\bigr(e^{-2\mathsf{W}_0(h_+,h_+)}+e^{-2\mathsf{W}_0(h_-,h_-)}\bigr)}{2}\,,
\end{align}
where $h_\pm = \pm f \mp 2E(f,\tilde\zeta)\tilde\zeta$. Using the fact that $\omega_0(e^{\ii\hat\phi(h_\pm)}) = e^{-2\mathsf{W}_0(h_\pm,h_\pm)}$, it follows that
\begin{align}
    \mathsf{W}_0(h_\pm,h_\pm ) &= \mathsf{W}_0(f,f) + 4E(f,\tilde{\zeta})^2\mathsf{W}_0(\tilde{\zeta},\tilde\zeta) \notag\\
    &\hspace{1cm} -  2E(f,\tilde\zeta)\text{Re}(\mathsf{W}_0(f,\tilde{\zeta})) \notag\\
    &\equiv \mathsf{W}_\zeta(f,f)\,.
\end{align}
This agrees with the computation in Eq.~\eqref{eq: squeezed-wightman} and it shows that indeed the local noise of squeezed states can be interpreted as a vacuum noise of a squeezed smearing function via replacement $f\to h_\pm$. The same approach can be used for the other coefficients involving $\omega_\zeta(S_fC_f)$ and $\omega_\zeta(S_f^2)$: we get
\begin{align}
    \omega_\zeta(S_f^2) &= \frac{1-\bigr(e^{-2\mathsf{W}_0(h_+,h_+)}+e^{-2\mathsf{W}_0(h_-,h_-)}\bigr)}{2}\,,
\end{align}
and $\omega_\zeta(S_fC_f) = 0$ since $\omega_\zeta$ is quasifree. The question of whether squeezing can reduce noise from field fluctuations is equivalent to the question of whether $h_\pm$ satisfies
\begin{align}
    \mathsf{W}_0(h_\pm,h_\pm) \leq \mathsf{W}_0(f,f)\,.
\end{align}
which can be checked by direct computation in specific examples such as the Minkowski spacetime.

In principle, we can generalize this to arbitrary Gaussian states obtainable from the vacuum state via a sequence of displacement and squeezing operations by making use of the sort of computations done in Eq.~\eqref{eq: displacement-sequence} and \eqref{eq: deformed-coherent}, hence our preceding calculations are very general.

\section{Application: quantum entropy of the field after interaction}
\label{sec: extra-stuff}

In this section we provide one application of the non-perturbative formalism, namely the computation of quantum entropies after a simple-generated interaction with a qubit detector. We will in fact see that it is possible to compute the field entropy exactly with the aid of \textit{replica trick} that is not based on path integral formalism.


As shown in Section~\ref{sec: channels}, after interaction with the detector we know that the field state becomes mixed (which we denoted by $\omega'$ in Section~\ref{sec: complementary}): this is of course not surprising since we have an joint interacting system and the subsystems do get entangled in general after the interaction\footnote{This will not be the case if we choose the detector to be in a state that is a fixed point of the channel, i.e., initial states that commute with the monopole operator $\hat{O}$. }. However, while the von Neumann entropy can be evaluated easily for the qubit detector, it is a completely different story for the field\footnote{This is so even with the fact that the GNS representation of the algebra $\pi_\omega(\W(\M))$ is a Type I von Neumann algebra \cite{fewster2019algebraic}.} due to the infinite-dimensionality of the Hilbert space of the field and the continuum of field modes (unlike finitely many harmonic oscillators).  

More precisely, if $\omega$ is the initial state then $\hat{\rho}_{\omega'}$ is the density matrix in the GNS representation of $\omega$ and we have
\begin{align}
    S(\hat{\rho}_{\omega'}) &= -\int_{\text{spec}(\hat{\rho}_{\omega'})}\hspace{-0.75cm} \dd\mu_{\rho'}(\lambda)\, \lambda\log_2\lambda \,,
    \label{eq: von-neumann-continuum}
\end{align}
where $\mu_{\rho'}$ defines a projective-valued measure associated to $\hat{\rho}_{\omega'}$. This expression is essentially the infinite-dimensional generalization of spectral decomposition evaluated on the spectrum of $\hat{\rho}_{\omega'}$ (see \cite{Hall2013quantum} for more details). There is no problem with this formula per se, but most of the time we do not have a good control over how to evaluate such an integral. The situation is worse if the algebra of observables we consider is a local algebra $\W(\mathcal{O})$ where $\mathcal{O}$ is some open subset of $\M$, since it gives rise to a  Type III algebra where von Neumann entropy simply does not exist (see, e.g., \cite{hollands2017entanglement} for details).  

\subsection{R\'enyi entropy and replica trick for von Neumann entropy}

First, let us demonstrate how to calculate the \textit{R\'enyi entropy} \cite{renyi1961measures} associated to the field state after backreaction from the qubit detector despite not having the explicit spectral decomposition of $\omega'$ in the sense of Eq.~\eqref{eq: von-neumann-continuum}. 
The quantum R\'enyi entropy of order $\alpha$ (hereafter $\alpha$-R\'enyi entropy) is defined to be 
\begin{align}
    S_\alpha(\hat{\rho}) = -\frac{1}{\alpha-1}\log_2 \tr(\hat{\rho}^\alpha)\,,
    \label{eq: Renyi-entropy}
\end{align}
where $\alpha\in [0,1)\cup (1,\infty)$. The limit $\alpha\to 1$ is the von Neumann entropy. Usually the quantum R\'enyi entropy is motivated and defined in terms of \textit{$\alpha$-R\'enyi divergence}, which generalizes the concept of relative entropy between two states. There are countless important applications of R\'enyi divergence, with several generalizations and operational implications (see, e.g., \cite{Wilde2013textbook} and refs therein). The special case where $\alpha=2$ is naturally identified with purity of a state\footnote{2-R\'enyi entropy is also distinguished by the fact that it is measurable experimentally without state tomography, i.e., computable without directly knowing the state of the system 
\cite{islam2015measuring}.}, which is also a measure of entanglement if the joint state is initially pure.

For simplicity, we first restrict our attention to the case when the field is in the quasifree state and the detector's initial state $\rod$ satisfies $\braket{\hat{O}}=0$. In terms of the density matrix in the GNS representation of $\omega$, we have
\begin{align}
    \hat{\rho}_{\omega'} &= C_{f}\ketbra{\Omega_\omega}{\Omega_\omega}C_{f} + S_{f}\ketbra{\Omega_\omega}{\Omega_\omega}S_{f}\,.
\end{align}
It follows that
\begin{align}
    \tr(\hat{\rho}_{\omega'}^2) &=\tr(\hat{\rho}_\omega C_{f}^2)^2+\tr(\hat{\rho}_\omega S_{f}^2)^2 + 2\tr(\hat{\rho}_\omega C_{f}S_{f})^2\notag\\
    &= \omega(C_{f}^2)^2+\omega(S_{f}^2)^2 + 2\omega(C_{f}S_{f})^2\notag\\
    &= \frac{1}{2}(1+\nu_{f}^2)\leq 1\,.
    \label{eq: purity}
\end{align}
In the third equality we have used the quasifree property that sets $\omega(S_fC_F)=0$. Hence the 2-R\'enyi entropy after interaction can be written in terms of smeared Wightman function
\begin{align}
    S_2(\omega') \equiv S_2(\hat{\rho}_{\omega'})&=  1-\log_2(1+\nu_f^2)\,,
    \label{eq: renyi-closed-form}
\end{align}
where $\nu_f=e^{-2\mathsf{W}({f},{f})}$ is the local noise factor. In effect, we have succeeded in computing the \textit{purity} of a quantum field state after a non-trivial interaction with a qubit detector.

How does Eq.~\eqref{eq: renyi-closed-form} compare to the entropy of the qubit detector? Suppose we take $\rod=a\ketbra{e}{e}+(1-a)\ketbra{g}{g}$ and $\hat{O}=\hat{\sigma}^x$ so that it fulfils $\braket{\hat{O}}=0$. Then the 2-R\'enyi entropy for the detector is given by
\begin{align}
    S_2(\Phi(\rod)) &= 1-\log_2\left(1+(1-2 a)^2\nu_f^2\right)\notag\\
    &\geq S_2(\omega')\,,
\end{align}
with equality only if $a=0,1$ (i.e., $\rod$ is pure). This shows that in general the detector's 2-R\'enyi entropy (purity) is bounded below by the field for arbitrary quasifree states subject to $\braket{\hat{O}}=0$.


The fact that we are only taking powers of $\hat{\rho}_{\omega'}^n$ for the computation of quantum R\'enyi entropy suggests that we might be able to actually calculate the resulting \textit{von Neumann entropy} of the field state after the interaction exactly, even without the knowledge of the spectrum. Indeed, we will now show that this can be achieved using the so-called \textit{replica trick} in QFT, and we do this without invoking the path integral representation of the field. The idea is to calculate, formally, the expression for the von Neumann entropy in terms of the R\'enyi entropy:
\begin{align}
    S(\hat{\rho}_{\omega'}) &= -\frac{\partial}{\partial n}\log \tr\hat{\rho}^n_{\omega'}\Bigr|_{n=1}\equiv \lim_{n\to 1}\frac{\tr(\hat{\rho}_{\omega'}^n)-1}{1-n}\,.
    \label{eq: von-Neumann}
\end{align}
Here we use natural logarithm for convenience and we can convert to base-2 logarithm at the end. 

 For clarity, let us consider the initially vacuum state of the field $\ket{0}$ which is also quasifree. Observe that we can recast the action of $C_f,S_f$ as producing a ``cat state'', i.e., superposition of coherent states $\ket{\pm \alpha_f}=\hat{D}(\pm \alpha_f)\ket{0}$ where $\alpha_f$ is the coherent amplitude associated to the spacetime smearing function $f\in \CS$ of the detector (\textit{c.f.} Section~\ref{sec: gaussian-operations}). We can write
\begin{equation}
\begin{aligned}
    \ket{C_f} \coloneqq C_f\ket{0} &\equiv \frac{1}{2}(\ket{\alpha_f}+\ket{-\alpha_f})\,,\\
    \ket{S_f} \coloneqq S_f\ket{0} &\equiv \frac{1}{2\ii }(\ket{\alpha_f}-\ket{-\alpha_f})\,.
\end{aligned}
\end{equation}
What is nice is that since $\braket{\pm\alpha_f|\pm\alpha_f}=1$ and from the Weyl algebra we get
\begin{align}
    \braket{-\alpha_f|\alpha_f} = \omega_0(\hat{D}(2\alpha_f)) = e^{-2\mathsf{W}_0(f,f)}\equiv \nu_f\,,
\end{align}
it follows that $\braket{C_f|S_f}=0$ --- that is, the two cat states are in fact orthogonal! Next, for convenience let us define $p_\pm = \frac{1 \pm \nu_f}{2} \in (0,1]$. It is now straightforward to calculate integral powers of $\hat{\rho}_{\omega'}$:
\begin{align}
    \hat{\rho}_{\omega'}^n  &= p_+^{n-1}\ketbra{C_f}{C_f} + p_-^{n-1}\ketbra{S_f}{S_f} 
    \label{eq: powers-density-matrix}
\end{align}
which gives
\begin{align}
    \tr(\hat{\rho}_{\omega'}^n) &= p_+^n+p_-^{n}\,.
    \label{eq: trace-renyi}
\end{align}
Finally, using the replica formula \eqref{eq: von-Neumann}, it follows immediately that the von Neumann entropy is 
\begin{align}
    S(\hat{\rho}_{\omega'}) &= -p_+\log_2 p_+ - p_-\log_2 p_-\,,
    \label{eq: von-neumann-vacuum}
\end{align}
which is nothing but the binary Shannon entropy with discrete probability distribution $\{p_\pm\}$. Note that this von Neumann entropy is explicitly calculable: all we need to know is the value of $\nu_f$, which only depends on the symmetrically smeared Wightman two-point function $\mathsf{W}(f,f)$. 

We have used the replica trick in the above because we wanted to get the von Neumann entropy by direct computation without knowing the explicit spectrum of the field state (which is difficult). That said, in hindsight there is a nice way to rewrite the calculations above in a manner that is analogous to the spectral decomposition in finite dimensions because of the orthogonality of the cat states: if this is possible then we could even avoid the replica trick altogether. Observe that since $\braket{C_f|S_f}=0$, we can think of $\hat{\rho}_{\omega'}$ as essentially being in a spectral decomposition over two orthogonal subspaces:
\begin{equation}
\begin{aligned}
    \hat{\rho}_{\omega'} &= p_+ \ketbra{+}{+} + p_-\ketbra{-}{-}\,, 
    \label{eq: infinite-projector-rep}
\end{aligned}
\end{equation}
where we define two vector states
\begin{align}
    \ket{+}\coloneqq p_+^{-1/2}\ket{C_f}\,,\quad \ket{-}\coloneqq p_-^{-1/2}\ket{S_f}\,,
\end{align}
which are well-defined since $p_\pm\geq 1/2$. These states give rise to two orthogonal projectors $\ketbra{\pm}{\pm}$ and these have the nice property that we can use the functional calculus on linear operators: using $f(x) = x^n$, we see that
\begin{align}
    f(\hat{\rho}_{\omega'}) &= p_+^n \ketbra{+}{+} + p_-^n \ketbra{-}{-}
\end{align}
which immediately yields Eq.~\eqref{eq: trace-renyi} by evaluating $\tr f(\hat{\rho}_{\omega'})$. Consequently, the von Neumann entropy is indeed the binary Shannon entropy with discrete probability distribution $\{p_\pm\}$ associated to the projectors $\ketbra{\pm}{\pm}$. Since the projectors are built from the cat states, strictly speaking these projectors can have infinite rank (as $S_f,C_f$ can have uncountable spectrum). 

How does the von Neumann entropy \eqref{eq: von-neumann-vacuum} compare with the von Neumann entropy for  the detector? Using the same initial detector state $\rod=a\ketbra{e}{e}+(1-a)\ketbra{g}{g}$ and $\hat{O}=\hat{\sigma}^x$ as before, we see that 
\begin{align}
    S(\Phi(\rod)) &= -(p_+ - a\nu_f)\log(p_+-a\nu_f)\notag\\
    &\hspace{0.4cm}  -\!(p_- + a\nu_f)\log(p_-+a\nu_f)\notag\\
    &\geq S(\hat{\rho}_{\omega'})\,,
\end{align}
with equality achieved only when $a=0,1$ (i.e., when $\rod$ is pure). Therefore, similar to 2-R\'enyi entropy we see that in general the von Neumann entropies are not equal and here we have a situation where we can perform exact computation of the quantum entropies for both the detector and the field after the interaction.

\subsection{Generalization to arbitrary Gaussian states}

We can actually avoid making restrictions about $\braket{\hat{O}}$ or restricting to quasifree states. The way to do this is to first rewrite the general field state in Eq.~\eqref{eq: field-state-general} after interaction as
\begin{align}
    \hat{\rho}_{\omega'} &= q_+\ketbra{+'}{+'} + q_-\ketbra{-'}{-'} \notag\\
    &\hspace{1cm} + \ii\braket{O}\sqrt{q_+q_-}\bigr(\ketbra{+'}{-'}-\ketbra{-'}{+'}\bigr)\,,
\end{align}
where $q_\pm = \frac{1}{2}(1\pm\Re\,\nu_f)$ and 
where now $\ket{\pm'}$ are now projectors associated to ``generalized'' cat states of the GNS vector $\ket{\Omega_\omega}$:
\begin{align}
    \ket{+} &= q_+^{-1/2}C_f\ket{\Omega_\omega}\,,\quad \ket{-} = q_-^{-1/2}S_f\ket{\Omega_\omega}\,.
\end{align}
Crucially, for generic Gaussian states the cat states are \textit{not} orthogonal, since
\begin{align}
    y \coloneqq \braket{+|-}=(q_+q_-)^{-1/2}\Im\,\nu_f\,.
\end{align}
For quasifree states we have $y=0$ and $q_\pm\to p_\pm$. 

The non-orthogonality of $\ket{\pm'}$ does not actually pose a problem. Following the similar strategy in Eq.~\eqref{eq: purity}, we can still calculate $\tr(\hat{\rho}_{\omega'}^n)$, which gives
\begin{equation}
    \begin{aligned}
    \tr(\hat{\rho}_{\omega'}^n) &= \Tilde{p}_+^n+\tilde{p}_-^n\,,
\end{aligned}
\end{equation}
where 
\begin{align}
    \Tilde{p}_\pm &\coloneqq \frac{1}{2}\bigg(1 \pm \sqrt{1-4(1-\langle\hat{O}\rangle^{2}) \left(q_+q_- -y^2\right)}\bigg)\,.
\end{align}
For $\braket{\hat{O}}=0$ and $y = 0$ we have $\tilde{p}_\pm\to p_\pm$ that we derived earlier. From this, the R\'enyi entropy and von Neumann entropy (via replica trick, for instance) can be computed to give
\begin{subequations}
\begin{align}
    S_n(\omega') &= -\frac{1}{n-1}\log_2(\Tilde{p}_+^n+\tilde{p}_-^n)\,,\\
    S(\omega') &= -\tilde{p}_+\log_2\Tilde{p}_+-\tilde{p}_-\log_2\Tilde{p}_-\,.
\end{align}
\end{subequations}
These are the results that apply for arbitrary Gaussian states of the field and also for arbitrary state of the detector where $\braket{\hat{O}}\neq 0$: thus the quantum entropies above form a three-parameter family indexed by $(\braket{\hat{O}},\Re\,\nu_f,\Im\,\nu_f)$. The quasifree state corresponds to the subfamily $\Im\,\nu_f = 0$.  

Overall, we have shown that for the non-perturbative interaction between a qubit and a quantum field prepared in an arbitrary Gaussian states, it is possible to calculate the quantum entropies of the field \textit{independently} of the the detector's entropy: in particular, unless the joint state is pure we have found that the entropy generated by the field is not the same as that of the detector. Furthermore, we showed this by directly computing the entropy of the field without performing infinite-dimensional spectral decomposition (where the computations are typically hard to control, if not impossible to extract the numbers directly). Our calculations work even for initially mixed states for either subsystem. The fact that they can be computed exactly using the conventional replica trick without knowing the detailed spectrum of the field is somewhat surprising. To the authors' knowledge, there is no simple path-integral representation to date for the large class of three-parameter family of states considered above, hence our calculations demonstrate the utility of the non-perturbative calculations to understand the field's output state, not just the qubit output state that is typically considered in the literature.

\section{Conclusion and outlook}

 In this work we collected and generalized several existing non-perturbative models for the interaction between a single two-level qubit detector and a relativistic quantum scalar field in arbitrary curved spacetimes, where the time evolution is given by simple-generated unitaries. We then extended the relativistic quantum channel associated to these non-perturbative models to include a very large class of Gaussian states of the quantum field. We showed how the results involving the non-vacuum Gaussian states can be rephrased in terms of those associated to the case when the field is in the vacuum state by embedding the displacement and squeezing operators into the Weyl algebra, effectively giving a ``Fourier transformed'' interpretation of the Gaussian operations in terms of the causal propagators in spacetime. Thus the extension for arbitrary Gaussian states of the field to those that are not necessarily quasifree turns out to be quite straightforward.
 
 One of the nice bonuses from our calculations is the fact that it is possible to show with minimal effort that for these simple-generated interactions, the R\'enyi entropy of the field state \textit{after interaction} with the detector can be calculated explicitly and independently of the calculations of the detectors' entropy. By using the replica trick, the von Neumann entropy for the field state can also be computed. These can be done without making any assumptions about the purity of the joint initial states of the detector and the field. Consequently, the non-perturbative models give us a three-parameter family of ``generalized cat states'' of the field whose entropies are finite and exactly computable. To the best of our knowledge, most of these states do not admit simple path integral representations, thus we believe our calculations are of independent interests.


There are several further extensions that we can consider following this work and we will briefly mention three of them that appear more immediately relevant. First, there are situations where one would like to think of multiple rapid-repeated interactions in delta-coupling model as being analogous to collision models \cite{Ciccarello2013collision,lorenzo2017nonmarkov,Grimmer2019zeno,Rybar2012simulate} and indeed this connection was studied for non-relativistic bosonic bath in \cite{Pipo2020spinboson}. The authors were able to frame the analysis in terms of Weyl relations of the canonical commutation relations, so we expect that relativistic generalization is straightforward and it is interesting to see if relativistic considerations have anything to say regarding CP-divisibility of the induced qubit channel. 

Second, the tractability of our calculations in this work suggests that at the very least, extending our results for two-qubits and three-qubit non-perturbative interactions may not be too difficult. For two-qubit systems, in particular, where two-party communication is most naturally set in, there has been  quite a few known results in the non-perturbative regimes in flat spacetimes (see, e.g., the thorough work in \cite{jonsson2018transmitting,Simidzija2020transmit,Simidzija2018nogo}), though this has changed recently to include curved backgrounds exploiting the sort of generalities we consider here \cite{Landulfo2016magnus1,tjoa2022channel,Tjoa2022fermi,Tjoa2022teleport,Landulfo2021cost}. The three-qubit system calculation has been only confined to entanglement and mutual information harvesting in flat space \cite{Gallock2021nonperturbative,mendez2022entanglement,avalos2022instant}, and there is also an example on sabotaging of correlations where they consider arbitrary number of detectors were considered in flat space) \cite{sahu2021sabotaging}. It is actually not difficult to show that there are ways to organize these calculations in the same spirit as this work in curved spacetimes. For every qubit introduced to the system, one can enlarge the family of field states whose quantum entropies can be computed exactly. Furthermore, because there are multiple parties involved in spacetime, it would be interesting to see how the entropies of the field behave as a function of causal relations between the detectors (i.e., the causal propagators). This is currently an ongoing investigation.

\section*{Acknowledgments}

E. T. thanks Eduardo Mart\'in-Mart\'inez for discussions regarding squeezing and coherent operations in QFT, and Jos\'e de Ramon Rivera for useful conversations regarding UDW detector formalism and its complications. E. T. also thanks Caroline Lima for useful discussions regarding the non-perturbative models. E. T. acknowledges that this work is made possible from the funding through both of his supervisors Robert B. Mann and Eduardo Mart\'in-Mart\'inez. This work is supported in part by the Natural Sciences and Engineering Research Council of Canada (NSERC).  The funding support from Eduardo Mart\'in-Mart\'inez is through the Ontario Early Research Award and NSERC Discovery program. The work has been performed at the Institute for Quantum Computing, University of Waterloo, which is supported by Innovation, Science and Economic Development Canada. University of Waterloo and Institute for Quantum Computing are situated on the Haldimand Tract, land that was promised to the Haudenosaunee of the Six Nations of the Grand River, and is within the territory of the Neutral, Anishnawbe, and Haudenosaunee peoples. E. T. would also like to thank both supervisors for providing the opportunity to explore some research ideas independently, through which this work becomes possible.

\bibliography{AQFT}

\end{document}